\documentstyle[12pt]{article}
\newcommand{\be}{\begin{equation}}
\newcommand{\en}{\end{equation}}
\newcommand{\fix}{\varphi (x,t)}
\newcommand{\fii}{\varphi }

\newcommand{\psip}{|\psi _p\rangle }
\newcommand{\psiz}{|\psi _z\rangle }
\newcommand{\psil}{\langle \psi }

\newcommand{\psipl}{\langle \psi _p}

\newcommand{\fil}{\langle \varphi }
\newcommand{\fip}{|\varphi _p\rangle }
\newcommand{\fipl}{\langle \varphi _p}

\newcommand{\Lo}{{\cal L}_0 }

\newcommand{\go}{\bar g_0}
\newcommand{\gi}{\bar g_1}
\newcommand{\D}{{\cal D} }

\newtheorem{defn}{Definition}
\newtheorem{rem}{Remark}
\newtheorem{lem}{Lemma}
\newtheorem{cor}{Corollary}
\newtheorem{teo}{Theorem}

\input{amssym.def}

\begin{document}

\hfill math-ph/xxyyzzz

\begin{center}
{\Large \bf Darboux transformation of coherent states\footnote {Talk at
the XI International Conference PROBLEMS OF QUANTUM FIELD THEORY, July
13 - 17, 1998, Dubna, Russia} \\}
\bigskip
{\large V. G. Bagrov, B. F. Samsonov and L. A. Shekoyan\\}
\date{}
\smallskip
{\it
Tomsk State University, Tomsk, Russia}
\end{center}
\begin{abstract}
It is proved that
the Darboux transformation of the system of coherent states of a
free particle leads to the states that may be treated as
coherent states of soliton-like potentials.
\end{abstract}
\bigskip {\Large \bf 1 Introduction }\\
\vspace{2mm}

\noindent
In quantum mechanics a nice recipe is known which permit us to
multiply exactly solvable quantum mechanical problems.
This method is
known under three different names. The first name is SUSI
QM \cite{wi}. It is originated from
quantum field theory.  The second name is the factorization method
proposed by Shr\"odinger \cite{sh}.  The third name is the Darboux
transformation method \cite{da}. This method is extensively
studied in soliton theory \cite{ma}.
It was Darboux who gave first the more clear formulation of it.
This method now is a part of a more general sheme known as the method of
tarnsformation operators [5].

In this paper Darboux transformation applied to
the free particle Shr\"odinger equation. It is proved that
transformed free particle coherent states are
coherent satets of soliton-like potentials. We use the
definition of coherent states given by Klauder \cite{kl}.

\bigskip
\hspace*{-6mm}{\Large \bf 2 Formalism}\\
\vspace{2mm}

\noindent
Let us know a general solution of the Schr\"odinger equation
\be \label {1}
(i\partial _t-h_0)\psi (x,t) =0,\quad
h_0=-\partial _x^2+V_0(x,t),\quad x\in [a,b].
\en
We want to find the solutions of another equation
\be \label {2}
(i\partial _t-h_0)\varphi (x,t) =0,\quad
h_1=-\partial _x^2+V_1(x,t),\quad x\in [a,b].
\en
The problem of the search for the solutions of the equation
(\ref{2}) can be reduced to the problem of looking for such an
operator $L$ that participates in the following intertwining
relation
\be \label {3}
L(i\partial _t-h_0)= (i\partial _t-h_1)L.
\en
Given the operator $L$ which satisfy this relation we easily find
function $\fix =L\psi (x,t)$.
It is clear that for an arbitrary Hamiltonian $h_1$ the problem
of looking for the operator $L$ is not more easy then the
problem of searching for the solutions of the Schr\"odinger
equation (\ref{2}).
It turns out to be that to make this method suitable for giving
useful information it is sufficient to restrict the operator
$L$ by any class of operators.
It is natural to look for the operator $L$ as a differential
operator. In this case $h_1$ can not be an arbitrary
Hamiltonian and equation (\ref{3}) becomes the equation for $L$
and $V_1$.

Let $L$ be an operator of the first degree in derivatives
$\partial _x$ and $\partial _t$. If we want to apply $L$ only
to the solutions of the equation (\ref{1}) we can replace
$\partial _t=-ih_0$ and it becomes of the second degree in
$\partial _x$. Therefor if we want that $L$ be of the first
degree in $\partial _x$ we should take the following form for
it
\be \label {4}
L=L_0(x,t)+L_1(x,t)\partial _x.
\en
The equation (\ref{3}) results then in the equation for the
functions $L_0$, $L_1$, and $V_1$. It is remarkable that this
system can be integrated and it has a vary nice and simple
solution \cite{phl}
\be \label {5}
L=L_1(t)(-u_x/u+\partial _x), \quad
L_1(t)=\exp [2\int dt{\rm Im}(\ln u)_{xx}],
\en
\begin{equation}
A=-(\ln \left| u\right| ^2)_{xx}.
\label{6}
\end{equation}
The function $u$ called the transformation fuction is a solution to the
initial
Schr\"o\-din\-ger equation subject to the condition
$(\ln u/\overline{u})_{xxx}=0 $.


\vspace{5mm}
\hspace*{-6mm}{\Large \bf 3 Coherent states}
\begin{defn}{\hspace{-0.5em}.} \label{def1}
Every system of states described by vectors
$\psiz $ is called the system of coherent states if the
following conditions are fulfilled{\rm :}
\newline
$({\rm i})$ $\psiz \in H_0$ where $H_0$ is a Hilbert space;
\newline
$({\rm ii})$ $z\in \D \subset \Bbb C;$
\newline
$({\rm iii})$ $\D $ is a domain endowed with a measure
$\mu (z,\bar z)$, $z,\bar z\in \D $ which is defined and finite
on a class of Borel sets of $\D $ and guaranties the following
resolution of the identity operator ${\Bbb I}$ on $H_0${\rm :}
\be \label {12}
\int _{\cal D}d\mu |\psi _z\rangle \langle \psi _z|={\Bbb I};
\en
$({\rm iv})$ $\forall z\in \D $, $\psiz $ belong to a domain of
definition of a Hamiltonian $h_0$ on $H_0$ and are
solutions to the Schr\"odinger equation
\be \label {13}
(i\partial _t-h_0)|\psi _z\rangle =0.
\en
\end{defn}
Consider the vectors $\fii _z(x,t)=L\psi _z(x,t)$. The question
that arises in this respect is the following. Whether the
vectors $\fii _z$ may be interpreted as coherent states of the
transformed system. It is clear that all the properties of
coherent states formulated in the Definition 1 are fulfilled
except may be for the resolution of the identity.

We shall consider now coherent states of soliton potential.

It is well known that the soliton potential
\be \label{14}
V_1(x)=-2a^2{\rm sech} ^2ax, \quad a>0
\en
can be obtained from the free particle Schr\"odinger equation,
$V_0(x,t)=0$ with the help of the Darboux transformation.
The Darboux transformation operator has the form
$L=-a{\rm th} ax +\partial _x$.
This operator together with its conjugate
$L^+=-a{\rm th} ax -\partial _x$ factorizes the free particle
Hamiltonian $h_0=L^+L-a^2$.
So, the operator $g_0=L^+L=h_0+a^2$ is strictly positive
definite and well defined in $H_0$ with the well-defined domain
of definition $D_0$.

The discrete basis \cite{mi} of the Hilbert space $H_0$ is defined with
the help of the raising and lowering operators
\be \label {15}
a=(i-t)\partial _x+ix/2,\quad a^+(i+t)\partial _x-ix/2,
\en
\be\label{16}
a^+\psi _n=\sqrt {n+1}\psi _{n+1},
\quad a\psi _n=\sqrt n \psi _{n-1}, \quad a\psi _0=0,\quad
\psi _n=\psi _n (x,t).
\en

The free particle coherent states may be defined as such
solutions of the equation (\ref{1}) that satisfy the equation
\be \label{17}
a\psi _z=z\psi _z,\quad z\in {\Bbb C} .
\en
The Fourier expansion of $\psi _z$ in terms of the basis
$\psi _n$ looks like as follows
\be \label{18}
\psi _z=\Phi \sum _n a_n z^n \psi _n, \quad
\Phi = \Phi (z,\bar z)=\exp (-z\bar z /2),\quad
a_n=(n!)^{-1/2}.
\en
These states satisfy the Definition 1 with the measure
$d\mu =dxdy/\pi $, $z=x+iy$.

The momentum operator $p_x$
and the Hamiltonian $h_0$
are expressed in terms of $a$ and $a^+$
\be
\label {19}
p_x=\frac 12 (a+a^+),\quad h_0=-p_x^2=\frac 14 (a+a^+)^2 .
\en

Consider the lineal ${\cal L}_0={\rm span}\{\psi _n\}$ which is
the space of the finite linear combinations of the functions
$\psi _n$.
Then $H_0=\bar {\cal L}_0$. (Bar over ${\cal L}_0$ means the
closure with respect to the norm generated by the
convential scalar product that we will label with the indice zero).

The operators $a$ and $a^+$ are completely defined on the
lineal $\Lo $ by the formulas (\ref{16}). Therefor we can
consider $\Lo $ as the initial domain of definition of $g_0$.
Since the operator $h_0=-\partial _x^2$ initially defined on
$\Lo $ has the deficiency indices equal zero it has the unique
self adjoint extension which coincides with its closure
$\bar h_0$, $\bar h_0=\bar h_0^+$ with the domain of definition
$D_0\subset H_0$.
The operator $g_0=h_0+a^2$ is essentially self adjoint as well,
$\bar g_0=\bar g_0^+$ and it has the same domain of definition
$D_0$. The spectrum of $\bar g_0$ is purely continuous.
The eigenfunctions $\psi _p(x,t)$ of the momentum operator
$p$ are the eigenfunctions of $\bar g_0$ as well
$\bar g_0\psi _p=N_p^2\psi _p$, $N_p^2=p^2+a^2$.

The Hamiltonian of the soliton potential,
$h_1=-\partial _x^2+V_1(x)$,
is essentially selfadjoint in $H_0$ and it has a mixed
spectrum.
It has a single discrete spectrum level
$E_{-1}=-a^2$ with the eigenfunction
\be \label{20}
\fii _{-1}(x,t)=(a/2)^{1/2}e^{-ia^2t}\cosh^{-1}(ax)
\en
Its continuous spectrum is the same that those of the
hamiltonian $h_0$. Let $\fii _p=\fii _p(x,t)$ be the continuous
spectrum eigenfunctions of $h_1$,
$h_1\fii _p=p^2\fii _p$.

It is easy to see that the action of the operator $L$ on the
basis functions $\psi _n$  is well defined and gives the
functions
\be \label {21}
\fii _n(x,t)=L\psi _n(x,t)
\en
which are solutions to the Schr\"odinger equation with the
soliton potential.

Let us consider the orthogonal decomposition
$L^2({\Bbb R})=L_0^2\oplus L_1^2$ where $L^2_0=\overline {\rm
span}\fii _{-1}$.  The functions $\fii _n$ are the basis functions in
the space $L_1^2$. The relation (\ref{21}) defines the action of $L$ in
the lineal $\Lo $. The natural question that arises at this
level is the following: What is the maximal domain of
definition of $L$.
Our analysis shows that the following lemma is valid.
\begin{lem}{\hspace{-0.5em}.}
The operator $L$ has such an extension $\bar L$ that it domain
of definition is $D^{\prime }_0$ and it domain of values is
$H_1$
where
$D^{\prime }_0=D_{\sqrt {g_0}}$ and $H_1=\bar {\cal L}_1$,
${\cal L}_1={\rm span}\{\fii _n\}$, $n=0,1,\ldots $,
and the closure is taken with respect to the norm generated by
the scalar product
$$\langle \fii _a|\fii _b\rangle _1\equiv
\langle \psi _a|\bar g_0|\psi _b\rangle _0,\quad
\psi _{a,b}\in {\cal L}_0, \quad \fii _{a,b}\in {\cal L}_1$$
\end{lem}
\begin{cor}{\hspace{-0.5em}.}
Every $\fii \in H_1$ may be presented in the form
$\fii =\bar L\psi $,
$\psi \in D^{\prime}_0$, $D^{\prime }_0\supset D_0$
\end{cor}

Let us define the operator $\bar L^+$ in the space $H_1$.
For this purpose let us consider the functions
$\psi \in D_0\subset D^{\prime }_0$ and for every $\fii =\bar
L\psi $,
$\psi \in D_0$ define $\bar L^+\fii \equiv \bar g_0\psi $.
Denote $D_1$ the domain of definition of $\bar L^+$.
Domain $D_1$ consists of all $\fii \in H_1$ of the form
$\fii =\bar L\psi $, $\psi \in D_0$. Domain $D_1$ is dense
in $H_1$. We have established the validity of the following lemmas
\begin{lem}{\hspace{-0.5em}.}
$\bar g_0=\bar L^+\bar L,\quad
\bar g_1=\bar h_1+a^2=\bar L \bar L^+$
\end{lem}
\begin {lem}{\hspace{-0.5em}.}
$\bar L^+$ is adjoint to $\bar L$ with respect to the scalar
products $\langle \cdot |\cdot \rangle _0$ and
$\langle \cdot |\cdot \rangle _1$
\end{lem}
\begin{lem}{\hspace{-0.5em}.}
$\bar L=\bar L^{++}$.
\end{lem}
\begin{cor}{\hspace{-0.5em}.}
The operator $\bar L$ is closed.
\end{cor}

The operator $\bar L$ has a natural extension to the continuous
spectrum eigenfunctions $\psi _p$ of the momentum operator and
$\bar L\psi _p=N_p\fii _p$, $N_p^2=p^2+a^2$, $p\in \Bbb R $.

The operator $\bar L^+$ is invertible in $H_1$. Introduce an
operator $M$ by the relation
\be \label {22}
M\bar L^+\fii =\fii , \quad \fii \in D_1,\quad
M=(\bar L^+)^{-1}
\en

\begin{lem}{\hspace{-0.5em}.}
The bases $\{\eta _n=M\psi _n\}$ and $\{\fii _n=L\psi _n\}$
are biorthogonal Riesz bases \cite{ni}.
\end{lem}

\begin {teo}{\hspace{-0.5em}.}
Operator $U=\bar L \go ^{-1/2}$ realizes the isometric mapping
of the domain $D^{\prime }_0$ onto $D_1$. Operator
$U^+=U^{-1}=\go ^{-1/2}\bar L^+$ realizes the inverse mapping.
Operators $U$ and $U^+$ have the following resolutions in terms
of the generalized eigenvectors $\psip $ and $\fip ${\rm :}
\be \label {UU}
U=\int dp \fip \psil _p|, \quad U^+=\int dp \psip \fil _p|.
\en
\end{teo}

\begin{cor}{\hspace{-0.5em}.}
>From {\rm (\ref{UU})} it follows the spectral representation
for
$\bar L$ and $\bar L^+$
$$\bar L=\int dp N_p\fip \psipl |, \quad
\bar L^+=\int dp N_p\psip \fipl |$$
and the similar representation for $M$ and $M^+$
$$M=\int dp N_p^{-1}\fip \psipl |, \quad
M^+=\int dp N_p^{-1}\psip \fipl |.$$
Operators $M$ and $M^+$ are bounded and factorize the
operators $\go ^{-1}$ and $\gi ^{-1}$:
$M^+M=\go ^{-1}$, $MM^+=\gi ^{-1}$.
\end{cor}

\begin{rem}{\hspace{-0.5em}.}
The representation $\bar L=U\go ^{1/2}$ is a canonical
representation of the closed operator $\bar L$ and
$M=U\go ^{-1/2}$ is the similar representation of the bounded
operator $M$. These
representations are
called polar factorizations as well.
\end{rem}

\begin{teo}{\hspace{-0.5em}.}
The states associated with the vectors
$\eta _z=M\psi _z=\Phi \sum _n a_nz^n\eta _n$
 are coherent states in the sense of the Definition 1. The
measure $d\mu _\eta = d\mu _\eta (z,\bar z)$
which realizes the resolution of the identity in terms of the
vectors $\eta _z$ gives a solution to the problem of moments in
the complex plane
$$a_n a_k \int d\mu _\eta |\Phi |^2 z^n \bar z^k=S_{nk}$$
and has the form
$$d\mu _\eta =\omega _\eta (x)dxdy,\quad
\omega _\eta (x)=\frac 1\pi (x^2+a^2-\frac 14),\quad
z=x+iy$$
\end{teo}

Consider now the vectors
$\fii _z=L\psi _z=\Phi \sum _n a_n z^n \fii _n$.

\begin{teo}{\hspace{-0.5em}.}
The states associated with the vectors $\fii _z$ satisfy all
the conditions of the Definition 1. The measure
$d\mu _\fii = d\mu _\fii (z,\bar z)$
which realizes the identity resolution in terms of the vectors
$\fii _z$ gives a solution to the problem of moments on the
complex plane
$$a_n a_k \int d\mu _\fii |\Phi |^2 z^n \bar z^k=S_{nk}^{-1}$$
and has the form
$d\mu _\fii =dyd\nu (x)$, $z=x+iy$. The measure $d\nu (x)$
is defined by its Fourier transform
$$d\tilde \nu (t)=\rho (t)dt, \quad
\rho (t)=\exp (t^2/8-a|t|)/(2\pi a)$$
\end{teo}
The work is partially supported by the grants of RFBR and Government
of Russia.

\end{document}